\documentclass{ws-procs9x6}

\begin{document}

\title{Momentum spectroscopy of phase fluctuations of an elongated Bose-Einstein condensate\footnote{\uppercase{T}his work is supported by \uppercase{CNRS},
 \uppercase{M}inist\`{e}re de la \uppercase{R}echerche,
 \uppercase{DGA}, \uppercase{EU} (\uppercase{C}old \uppercase{Q}uantum \uppercase{g}ases network;
 \uppercase{ACQUIRE} collaboration),
 \uppercase{INTAS}(project 01-0855)}}

\author{A. ASPECT, S. RICHARD, F. GERBIER, M. HUGBART, J. RETTER, J. H. THYWISSEN\footnote{\uppercase{P}resent adress:
\uppercase{D}epartment of \uppercase{P}hysics, 60 \uppercase{S}t
\uppercase{G}eorge \uppercase{S}treet, \uppercase{U}niversity of
\uppercase{T}oronto, \uppercase{T}oronto, \uppercase{ON, M5S 1A7},
\uppercase{C}anada }, and P. BOUYER}

\address{Laboratoire Charles Fabry, Institut d'Optique, \\
Bat. 503, Centre Universitaire \\
F91403, ORSAY, France\\
E-mail: alain.aspect@iota.u-psud.fr}


\maketitle

\section{Introduction: BEC beyond the ideal case}

Degenerate atomic Bose gases provide a remarkable testing ground
for the theory of dilute quantum fluids, allowing for an extensive
comparison of theory and experiment\cite{DalfovoRMP,nobelrmp}.

First studies on trapped dilute Bose gases have shown behaviors in
good agreement with the predictions for homogeneous ideal Bose
gases. With the refinement of experimental techniques, however, it
becomes possible to study deviations from the ideal case. We have
recently made quantitative studies of such deviations, for two
different phenomenon:
\begin{romanlist}
\item Phase fluctuations in an elongated quasi 1D condensate,
that lead to a coherence length smaller than the condensate axial
size\cite{Richardspectro}.
\item Shift of the critical temperature of condensation due to
interactions\cite{GerbierTc}.
\end{romanlist}

\noindent In the context of this series of Conferences On Laser
Spectroscopy, where we have so strong traditions, I will
concentrate on the first study, since we will see that these
measurements involve many methods of "Good Old Spectroscopy",
transposed from the world of Lasers and Photon Optics, to the
world of Bose-Einstein Condensates and Atom Optics.

\section{Phase fluctuations in a trapped elongated BEC}
Three-dimensional (3D) trapped condensates are predicted to have a
uniform phase, extending over the whole condensate, even at finite
temperature, and this has been confirmed
experimentally\cite{Phillips_coherence1,stenger99,Bloch}. In low
dimensional systems, however, phase fluctuations are predicted to
reduce the spatial coherence length (see
\cite{petrov1d,StoofGriffinCastin} and references therein). This
phenomenon also occurs in sufficiently anisotropic 3D samples,
where complete phase coherence across the axis (long dimension) is
established only below a temperature $T_{\rm{\phi}}$, that can be
much lower than the critical temperature $T_{\rm{c}}$
\cite{petrov3d}. In the range $T_\phi < T < T_{\rm c}$, the cloud
is a ``quasi-condensate'', whose incomplete phase coherence is due
to thermal excitations of 1D axial modes, with wavelengths larger
than its radial size. Phase fluctuations of quasi-condensates in
elongated traps have been observed first in Hannover by Dettmer
{\em et al.}\cite{dettmer01}, who measured the conversion, during
free expansion, of the phase fluctuations into ripples in the
density profile. Although the conversion dynamics is well
understood\cite{Ertmer3}, the measured amplitude of density
ripples was found smaller than expected by a factor of two. In
Amsterdam, Shvarchuck {\it{et al.}}\cite{shvarchuck02} have used a
``focusing" method to show that the coherence length is smaller
than the sample size during the formation of a condensate.

\section{Production and observation of an elongated BEC
in an iron-core electromagnet} Our experimental setup has been
described elsewhere\cite{interrupted}. Briefly, a Zeeman-slowed
atomic beam of $^{87}$Rb is trapped in a MOT, and after optical
pumping into the $F=1$ state is transferred to a magnetic
Ioffe-Pritchard trap created by an iron-core electromagnet. A new
design allows us to lower the bias field to a few Gauss and thus
to obtain very tight radial confinement\cite{these_vincent}. Final
radial and axial trap frequencies ($\omega_\perp$ and $\omega_z$)
are respectively 760 Hz and 5 Hz (\textit{i.e.} an aspect ratio of
$152$). We typically obtain needle-shaped condensates containing
around $5 \times 10^4$ atoms, with a typical half-length
$L\simeq130\,\mu$m and radius $R\simeq0.8\,\mu$m. Since the
chemical potential is only slightly larger than the radial
vibration quantum, (${\mu_{\rm TF}}/{\hbar \omega_\perp}\sim 4$),
we work at the crossover between the 3D and 1D Thomas-Fermi (TF)
regimes\cite{gorlitz01,schreck01,greiner01}.

To obtain the number of atoms in the condensate and in the thermal
cloud, and the temperature, we make an absorption image after a 20
ms time of flight. The number of atoms, obtained to within 20\%,
is calibrated from a measurement of the temperature of
condensation $T_\textrm{c}$. The temperature $T$ is extracted from
a two-component fit to the absorption images, yielding the
temperature of the thermal cloud fitted by an ideal Bose function
with zero chemical potential. In fact, the temperature obtained at
the end of the evaporation ramp can be chosen \textit{a priori} to
within 20\,nK by controlling the final trap depth (final rf
frequency, as compared to the one totally emptying the trap, with
is checked every five ramps) to a precision of 2\,kHz. This
provides a high reproducibility which is a real asset for these
experiments.

As Shvarchuck {\it{et al.}}\cite{shvarchuck02}, we observe strong
shape oscillations at the formation of the
condensate\cite{Fabrice_proceed}, despite a slow evaporation (less
than 100\,kHz$/$s) across $T_{\rm c}$. We then hold the condensate
for a  time of typically 7 seconds in the presence of an rf
shield, to damp the axial oscillations enough that they do not
affect the Bragg spectra (see below).

\section{Measurement of the spatial coherence function by Bragg spectroscopy}
The most direct way to measure the spatial coherence function
relies on atom interferometry. It turns out, however, that this
 method demands a high level of shot to shot stability of the
 interferometer\cite{hannovreinterfero},  and we have decided to
 first use a complementary method, that directly yields the
 momentum distribution (momentum spectrum) $\mathcal{P}(p_z)$,
 which is nothing else than  the
 Fourier transform of $C(s)\,$.

 This is analogous to the two complementary methods of
 ``Good Old Spectroscopy'', where one can either measure the
 temporal correlation function of the light (the so called
 ``Fourier-transform spectroscopy"), or directly obtain the
 frequency spectrum. It is well known that the second type of
 method is much easier to implement, because some physical
 phenomena
(as dispersion in dielectric media) or "simple" devices (as
 diffraction gratings) are able to separate frequency
 components whose weight can be directly measured.
 The methods yielding directly the spectrum were
 in fact the only ones that were
 used for centuries, until stable and reliable enough
 interferometers became available.

In our problem of Atom Optics, it turns out that we also have a
method that directly yields the momentum spectrum: this is ``Bragg
Spectroscopy"\cite{stenger99,davidson}. Our momentum distribution
measurement is based on four-photon velocity-selective Bragg
diffraction. In this process, atoms are extracted out of the
condensate by interaction with a moving standing wave, formed by
two counter-propagating laser beams with a relative detuning
$\delta$. In a wave picture, this can be interpreted as a (second
order) Bragg diffraction of the atomic matter waves off the
grating formed by the light standing wave, and it is therefore
resonant only for a given atomic de Broglie wavelength, or
equivalently for a given value of the atomic momentum. The number
of extracted atoms is therefore proportional to the density of
probability of that particular value of the momentum. Writing the
Bragg condition of diffraction off a thick grating, one finds that
the momentum component resonantly diffracted out of the condensate
depends on the velocity of the light standing wave, and therefore
on the detuning $\delta$, according to

\noindent
\begin{equation}
p_z=M(\delta-8\omega_{\rm{R}})/(2k_{\rm L}) \,, \label{eq:e2}
\end{equation}

\noindent with $\omega_{\rm{R}}=\hbar k_{\rm{L}}^2/(2M)$, $M$ the
atomic mass, and $k_{\rm L}=2\pi/\lambda$ ($\lambda=780.02$\,nm).
By varying the detuning $\delta$ between the counter-propagating
laser beams that make the moving standing wave, and measuring the
fraction of diffracted atoms \textit{vs.} $\delta$, one can build
the momentum distribution spectrum.

The stability of the differential frequency $\delta$ determines
the spectral resolution, and must be as good as possible. The
optical setup is as follows. A single laser beam is spatially
filtered by a fiber optic, separated into two arms with orthogonal
polarizations, frequency shifted by two independent 80\,MHz
acousto-optic modulators, and recombined. The modulators are
driven by two synthesizers stable to better than 1\,Hz over the
typical acquisition time of a spectrum. The overlapping,
recombined beams are then sent through the vacuum cell, parallel
(to within 1\,mrad) to the long axis of the trap, and
retro-reflected to obtain two standing waves with orthogonal
polarizations, moving in opposite directions.  To check the
differential frequency stability, we have measured the beat note
between the two counter-propagating beams forming a standing wave.
The average over ten beat notes had a half-width at half-maximum
(HWHM) of $216(10)$\,Hz for a 2\,ms pulse\cite{beat}.

\section{Axial Bragg spectrum of an elongated condensate}

The following experimental procedure is used to acquire a momentum
spectrum. At the end of forced evaporative cooling, the radio
frequency knife is held fixed for about 7\,s to allow the cloud to
relax to equilibrium (see above). The magnetic trap is then
switched off abruptly, in roughly 100\,$\mu$s, and the cloud
expands for 2\,ms before the Bragg lasers are pulsed on for 2\,ms.
The lasers are tuned 6.6\,GHz below resonance to avoid Rayleigh
scattering, and the laser intensities (about 2 mW/cm$^2$) are
adjusted to keep the diffraction efficiency below 20\,\%.

We take the Bragg spectrum after expansion rather than in the trap
to overcome two severe problems. First, in the trapped condensate,
the mean free path (about 10\,$\mu$m) is much smaller than its
axial size of  260\,$\mu$m, so that fast Bragg-diffracted atoms
would scatter against the cloud at
rest\cite{Chikkatur_collisions}. Second, the inhomogeneous mean
field broadening\cite{stenger99} would be of the order of 300\,Hz,
\textit{i.e.} larger than our instrumental resolution. By
contrast, after 2\,ms of free expansion, the peak density has
dropped by two orders of magnitude\cite{castin_kagan}, and both
effects become negligible. One may wonder whether this 2\,ms
expansion doesn't affect the momentum distribution we want to
study. In fact, the phase fluctuations do not significantly evolve
in 2\,ms, since the typical timescale for their complete
conversion into density ripples varies from 400\,ms to 15\,s for
the range of temperatures we explore\cite{Fabrice_theory}. Also,
the mean field energy is released almost entirely in the radial
direction, because of the large aspect ratio of the
trap\cite{castin_kagan}, and contributes only about $50$\,Hz of
Doppler broadening in the axial direction. The only perturbation
due to the trap release seems to be small axial velocity shifts
(around $100\,\mu$m/s) attributed to stray magnetic gradients that
merely displace the spectra centers.

After application of the Bragg pulses, we wait for a further
20\,ms to let the diffracted atoms separate from the parent
condensate, and take an absorption image. Diffraction efficiency
is defined as the fraction of atoms in the secondary cloud. We
repeat this complete sequence typically five times at each
detuning, and we average the diffraction efficiency for this
particular value of the detuning. Repeating this process at
various values of the detuning (typically 15), we plot the
diffraction efficiency \textit{vs.} $\delta$, and obtain an
``elementary spectrum''.

\begin{figure}[ht]
\centerline{\epsfxsize=3.9in\epsfbox{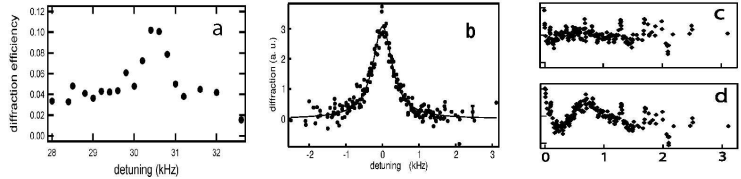}}
\caption{\textbf{Bragg spectrum of an elongated BEC} at
$T=261(13)$\,nK, corresponding to $T/T_\phi=20(2)$.\textbf{(a)}
Elementary spectrum (\textit{i.e.} diffraction efficiency
\textit{vs.} relative detuning of the Bragg lasers) corresponding
to 75 diffraction efficiency measurements made at the same
temperature, but at various detunings (see text). Measurements at
the same detuning (typically 5) are averaged. \textbf{(b)}
Averaged spectrum, obtained by averaging 12 (recentered and
rescaled, see text) elementary spectra. A typical statistical
error bar is shown. This spectrum is the superposition of 12
``elementary spectra'', as described in the text.  The solid line
is a Lorentzian fit, giving a half-width at half-maximum (HWHM) of
316(10)\,Hz. {\bf (c)} and {\bf (d)} show respectively the
(folded) residuals of a Lorentzian and of a Gaussian fit to the
above spectrum.   \label{spectres}}
\end{figure}

As shown on Fig.~\ref{spectres}a, an elementary spectrum at a
given temperature still shows a lot of noise. To average out this
noise, we have taken many elementary spectra at the same
temperature (between 10 and 40), but during different runs (on
different days). We have also varied the hold time (of about 7\,s,
see above) over a range of 125\,ms, to average on possible
residual shape oscillations. We then take all the spectra
corresponding to the same temperature, and we reduce them to the
same surface, background, and center, and superpose
them\cite{simon-these}.

Fig.~\ref{spectres}b shows the result of such a data processing,
at a temperature well above $T_\phi$, \textit{i.e.} for a case
where the broadening due to phase fluctuations is expected to
dominate over the instrumental resolution. The width (316(10)\,Hz,
HWHM) is definitely larger than the resolution (less than
216(10)\,Hz,HWHM). Moreover, it is clear by simple visual
inspection, and confirmed by examining the residuals of fits (see
Fig.~\ref{spectres}c an d) that the profile shape is definitely
closer to a Lorentzian than to a Gaussian. Such a shape is in
contrast to the gaussian-like profile expected for a pure
condensate\cite{stenger99,Zambelli}, and it is characteristic of
large 1D phase fluctuations\cite{Fabrice_theory}, which result in
a nearly exponential decay of the correlation function.

\section{Results. Comparison to theory}

\begin{figure}[ht]
\centerline{\epsfxsize=3.9in\epsfbox{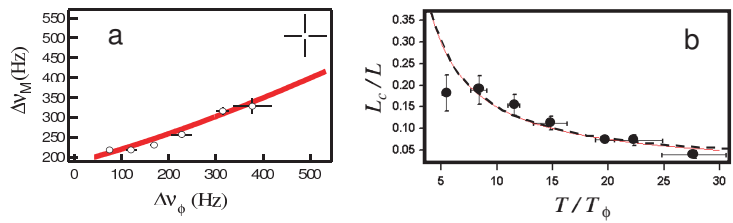}}
\caption{\textbf{Momentum distribution width and coherence length
for} $T/T_\phi >1$. \textbf{(a)} Half-widths at half-maximum
$\Delta\nu_{\rm M}$ of the experimental Bragg spectra versus the
parameter $\Delta\nu_\phi$ (see text).  The solid line is a fit
assuming a Voigt profile with a constant Gaussian resolution
function. \textbf{(b)} Coherence length $L_\textrm{c}$(divided by
the condensate half size $ L$) \textit{vs.} temperature (divided
by $T_\phi$). The coherence length is obtained after deconvolution
from the resolution function. \label{largeur_vs_temp}}
\end{figure}

Bragg spectra have been taken at various temperatures between
$T_\phi$ and $T_{\rm c}$. Using a Lorentzian fit, we extract a
measured half-width $\Delta\nu_\textrm{M}$ for each temperature,
and plot it (Figure~\ref{largeur_vs_temp}a) \textit{vs.}
$\Delta\nu_\phi$, a parameter convenient to compare to the theory.
That parameter can be directly expressed as a function of the
number of atoms and the temperature, and  theory predicts that the
ideal spectrum half-width should be proportional to
$\Delta\nu_\phi$, with a multiplying factor $\alpha$, depending on
the density profile, of the order of 1 ($\alpha=0.67$ in our
case)\cite{Fabrice_theory,Richardspectro}. In fact, since our
spectral resolution is limited, the measured spectrum width
$\Delta\nu_\textrm{M}$ results from a convolution of the ideal
Lorentzian profile by the resolution function, that we assume to
be a Gaussian, of half-width $w_{\rm G}$. The measured spectrum is
then expected to be a Voigt profile, whose shape can hardly be
distinguished from a Lorentzian in our range of parameters, but
with a width (HWHM) $\alpha\Delta\nu_\phi/2+\sqrt{w_{\rm
G}^2+(\alpha\Delta\nu_\phi)^2/4}$. We use that expression to fit
the data of Fig.~\ref{largeur_vs_temp}a, taking $\alpha$ and
$w_{\rm G}$ as free parameters. We find $w_{\rm G}=176(6)$\,Hz,
and $\alpha= 0.64(5)(5)$. The first uncertainty quoted for
$\alpha$ is the standard deviation of the fit value. The second
results from calibration uncertainties on the magnification of the
imaging system and on the total atom number, which do not affect
$w_{\rm G}$. We note first that the fitted value of $w_{\rm G}$ is
slightly smaller than $216$\,Hz, as it should be\cite{beat}. The
agreement of the measured value of $\alpha$ with the theoretical
value 0.67, to within the 15\,\% experimental uncertainty,
confirms quantitatively the temperature dependence of the momentum
width predicted in Ref.~\cite{petrov3d}.

\section*{Conclusion}
We have shown that the momentum spectrum of an elongated
condensate at a temperature smaller than $T_\textrm{c}$ but larger
than $T_\phi$ has a Lorentzian shape, in contrast with the
Gaussian shape of  a "normal" condensate. That shape, as well as
the measured broadening, agree quantitatively with the predictions
for a phase fluctuating condensate. The coherence lengths
corresponding to the spectrum widths
 are  smaller than the condensate
axial size (Fig.~\ref{largeur_vs_temp}b). It would be interesting
to also measure the coherence length for temperatures approaching
$T\phi$, to study how coherence develops over the whole
condensate. Since this corresponds to smaller and smaller momentum
 widths, the method presented here is not well adapted, and an
 interferometric measurement directly yielding the spatial correlation
 function at large separations would be a method of choice. Such
 measurements require a large path difference of the
 interferometer, and the stability of the interferometer is a crucial
 issue. We are currently working in that direction.

%
%
%
%

\end{document}